\documentclass[twocolumn,twocolappendix]{aastex63}
\pdfoutput=1 %for arXiv submission
\usepackage{amsmath,amstext}
\usepackage[T1]{fontenc}
\usepackage[figure,figure*]{hypcap}
\usepackage [english]{babel}
\usepackage [autostyle, english = american]{csquotes}
\usepackage {booktabs}
\usepackage{subfigure}
\usepackage{amssymb}
\usepackage{multirow}
\usepackage{gensymb}
\usepackage{amsmath}
\usepackage{hyperref}

\include{cmds}
\shortauthors{Guo et al.}

\begin{document}

\title{Searching for Barium Stars from the LAMOST Spectra Using the
Machine Learning Method: I}

\author{Fengyue Guo}
\author{Zhongding Cheng}
\author{Xiaoming Kong}
\email{xmkong@sdu.edu.cn}
\author{Yatao Zhang}
\affiliation{School of Mechanical, Electrical \& Information Engineering, Shandong University, Weihai, 264209, Shandong, People's Republic of China}
\author{Yude Bu}
\affiliation{School of Mathematics and Statistics, Shandong University, Weihai, 264209, Shandong, People's Republic of China}
\author{Zhenping Yi}

\author{Bing Du}
\affiliation{CAS Key Laboratory of Optical Astronomy, National Astronomical Observatories, Beijing 100101, People's Republic of China}

\author{Jingchang Pan}
\affiliation{School of Mechanical, Electrical \& Information Engineering, Shandong University, Weihai, 264209, Shandong, People's Republic of China}

%\accepted{\it 2022 11 10}
%\published{\it 2022 * *}
%\submitjournal{\it The Astronomical Journal}

%--------------------------------------------------------------------------------------------

\begin{abstract}

Barium stars are chemically peculiar stars that exhibit enhancement
of s-process elements. Chemical abundance analysis of barium stars
can provide crucial clues for the study of the chemical
evolution of the Galaxy. The Large Sky Area Multi-Object Fiber
Spectroscopic Telescope (LAMOST) has released more than 6 million
low-resolution spectra of FGK-type stars by Data Release 9 (DR9),
which can significantly increase the sample size of barium stars. In this paper, we used machine learning algorithms to search
for barium stars from low-resolution spectra of LAMOST. We have
applied the Light Gradient Boosting Machine (LGBM) algorithm to
build classifiers of barium stars based on different features, and
build predictors for determining [Ba/Fe] and [Sr/Fe] of barium
candidates. The classification with features in the whole spectrum
performs best: for the sample with strontium enhancement, Precision
= 97.81\%, and Recall = 96.05\%; for the sample with barium
enhancement, Precision = 96.03\% and Recall = 97.70\%. In
prediction, [Ba/Fe] estimated from Ba\,{\sc ii} line at 4554 {\AA}
has smaller dispersion than that from Ba\,{\sc ii} line at 4934
{\AA}: MAE$_{4554 \AA}$ = 0.07, $\sigma_{4554 \AA}$ = 0.12. [Sr/Fe]
estimated from Sr\,{\sc ii} line at 4077 {\AA} performs better than
that from Sr\,{\sc ii} line at 4215{\AA}: MAE$_{4077 \AA}$ = 0.09,
$\sigma_{4077 \AA}$ = 0.16. A comparison of the LGBM and other
popular algorithms shows that LGBM is accurate and efficient in
classifying barium stars. This work demonstrated that machine
learning can be used as an effective means to identify chemically
peculiar stars and determine their elemental abundance.

\end{abstract}

% ---------------------------------------------------------------------------
\section{Introduction}
% ---------------------------------------------------------------------------

Barium (Ba) stars are a type of G-K stars which were first
discovered by \cite{1951ApJ...114..473B}. They show strong
absorption lines from carbon and slow neutron capture process
(s-process) elements in their spectra, especially Ba\,{\sc ii} at
4554 {\AA} and Sr\,{\sc ii} at 4077 {\AA}. Ba stars have been
believed to originate from the evolutionary channel of binary stars
\citep{1983ApJ...268..264M}. The current research shows that the
enrichment of the s-process on the barium surface is likely to come
from the pollution caused by the mass transfer of the companion star
on the evolution stage of Asymptotic Giant Branch \citep{1988A&A...205..155B,1995MNRAS.277.1443H,1998A&A...332..877J,2011AJ....141..160G,2018MNRAS.474.2129K}.
Detailed chemical composition of barium stars can provide clues to
their origin, properties, and contribution to the Galactic chemical
enrichment.

Since barium stars were recognized in 1951, the observational and
theoretical studies on them have never stopped.
\cite{1972AJ.....77..384M} provided a large homogeneous sample of
241 barium stars which included "certain" and "marginal" barium
candidates. Then \cite{1991AJ....101.2229L} built a catalog with 389
barium stars, and determined the barium intensities (from 1 to 5)
and spectral classifications by analyzing image tube spectra and
photometric observation data. Other studies on such stars,
especially abundance analysis based on high-resolution spectroscopy,
are mostly based on a few or a dozen samples \citep{1979Tomkin,
1981Sneden, 1984Smith, 1997Porto,2005Pereira, 2003Liang,
2006A&A...454..895A,2007Gray, 2008Pomp, Pereira2011,
2016RAA....16...19Y, Merle2016, 2018Karinkuzhi}. \cite{Castro2016}
presented a homogeneous analysis of photospheric abundances based on
high-resolution spectroscopy of a sample of 182 barium stars and
candidates. All stars analyzed in their work were selected from
previous literature \citep{1972AJ.....77..384M, 1981Bidelman,
1991AJ....101.2229L}, and 13 out of 182 samples proved to be normal
stars because of their low mean s-process element abundances.
Therefore, the number of barium stars is still small and has not
been effectively extended in the past few decades. Some stars have
been investigated more than once, and a substantial fraction of them
was kicked out from the barium catalog \citep{1987Smith,
Smiljanic2007}

In view of the above-mentioned situation, a large sample of
barium stars with barium abundance is very useful in order to better
understand their origin and properties. Large-field spectroscopic
Surveys like the Large Sky Area Multi-Object Fiber Spectroscopic
Telescope (LAMOST), give us the chance to increase the sample size.
\cite{2018ApJS..234...31L}(hereafter L18) found 719 barium stars
with strong spectral lines in Ba\,{\sc ii} at 4554 and Sr\,{\sc ii}
at 4077 {\AA} when they identified carbon stars from LAMOST DR4 by
using a machine-learning method.
\cite{2019MNRAS.490.2219N}(hereafter N19) reported 895 (out of
454,180 giants) barium giant candidates which were classified into
49 Ba-only, 659 Sr-only and 49 both Sr- and Ba- enhancement stars
from low-resolution spectra of LAMOST DR2. They predicted the
stellar parameters and abundances ($T_\mathrm{eff}$, $\log g$,
$\mathrm{[Fe/H]}$, $\mathrm{[\alpha/M]}$) for their 454,180 giant
samples based on a training sample which transfers labels from
APOGEE to LAMOST by using the machine-learning method. Then, they
identified the s-process-rich candidates by comparing the strengths
of the Sr\,{\sc ii} (4077 and 4215 {\AA}) and Ba\,{\sc ii} lines
(4554 and 4934 {\AA}) between the observed flux and the Cannon model
\citep{2015Ness}. Finally, they estimated $\mathrm{[Ba/Fe]}$ and
$\mathrm{[Sr/Fe]}$ abundance ratios for all s-process-rich
candidates by spectrum synthesis.

To perform verification on the study of N19, \cite{2021Karinkuzhi}
selected 15 of the brightest targets from the s-process-rich
candidates provided by N19 and carried on high-resolution spectral
observations on them. The s-process element abundance analysis shows
that about 68\% of Sr-only and 100\% of Ba-only stars from the study
of N19 are true barium stars. The reason why Sr-only candidates were
misclassified is that Sr\,{\sc ii} lines at 4077 and 4215 {\AA} are
easily saturated, especially for the 4215 {\AA} line which is
affected by the strong CN bandhead at $\lambda$ 4216 {\AA}. The
study of \cite{2021Karinkuzhi} shows that three no-s stars which are
considered as Sr-only by N19 are those that are N-rich. As a matter
of fact, Sr and Ba element abundance estimation based on a
low-resolution spectrum is not easy. Other lines like Sr\,{\sc ii}
at 4607 {\AA}, 4811 {\AA} and 7070 {\AA} which are used in
\cite{2021Karinkuzhi} are not prominent enough and are easily
affected or covered by their nearby lines in low-resolution spectra.
Owing to the above reasons, the barium candidates provided by N19
are still valuable samples for the study of barium stars, especially
the samples with barium enrichment.

This study aims to explore the application of machine-learning
methods to search for barium stars based on the sample provided by
N19 and L18. A large number of machine-learning algorithms have been
used in the analysis of astronomical data and they perform well in
the estimation of atmospheric parameters and element abundance. In
this paper, we propose two kinds of models. One is for searching
barium enrichment candidates, and the other is for estimating the
$\mathrm{[Ba/Fe]}$ and $\mathrm{[Sr/Fe]}$ abundance ratios of the
candidates. Compared with various intricate algorithms, the Light
Gradient Boosting Machine (LGBM) algorithm performs best in terms of
precision and recall for the identification of Sr-enhanced
candidates and Ba-enhanced candidates, especially in the features of
the whole spectrum. The results obtained by the abundance prediction
model are in good agreement with the label.

This paper is organized as follows: Section \ref{sec:data} describes
the data set used in our experiment and the data preprocessing process. In Section \ref{sec:method}, we
introduce the LGBM and SVM algorithm. In Section
\ref{sec:experiment}, we show the construction and performance of
the proposed model including the classifier and predictor. Finally,
a short discussion and conclusion are given in Section
\ref{sec:conclusion}.

% ---------------------------------------------------------------------------
\section{Data}
\label{sec:data}
% ---------------------------------------------------------------------------

The LAMOST is a reflecting Schmidt telescope with a 3.6-4.9 m
effective aperture and 5{\degree} field of view, which is located in
the northeast of Beijing, China \citep{2012RAA....12.1197C}. LAMOST
can simultaneously observe up to 4000 objects in a single exposure
by the distributive parallel-controllable fiber positioning
technique in the latest release. LAMOST DR9 has released
10,907,516 spectra of stars with spectral resolutions of
R$\sim$1800, and wavelength coverage ranging from 370 to 900 nm.
\begin{figure}[htb] \centering \includegraphics[width=9cm]{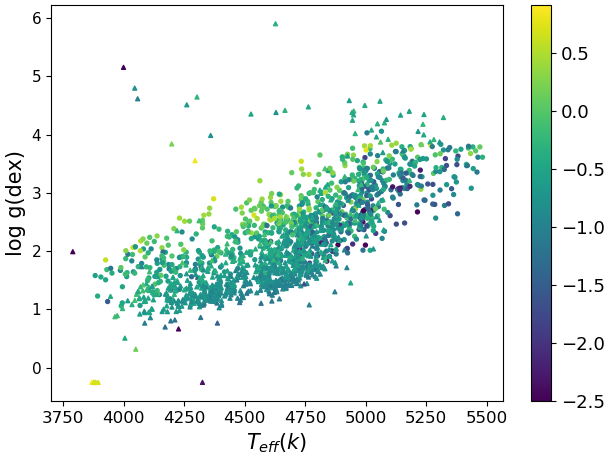}
\caption{Distribution of barium stars adopted as positive
samples in the plane of $T_\mathrm{eff}$-$\log g$, the colors
indicate the [Fe/H] of stars. The triangle symbols indicate
the barium star sample provided by L18, and the dot symbols indicate
samples provided by N19.}\label{fig:fig3} \end{figure}

In order to train and test the classifier, we collected
known samples of barium stars and non-barium stars. The data set of
barium stars used in this work has two parts. The first part
consists of 867 s-process-rich stars. These stars were obtained by
cross-matching LAMOST DR9 catalog with the catalog provided by N19,
including 48 Ba-only, 642 Sr-only, and 177 both Sr- and Ba-
enhancement candidates. The second part consists of 810 barium star
spectra. After cross-matching LAMOST DR9 catalog with 719 barium
stars provided by L18, we obtained 577 barium stars with 907
spectra. Then removing repeated spectra with signal-to-noise ratio
(S/N) \textless 30 pixel$^-1$ in 907 spectra, we finally
obtained 810 spectra.

In addition, the spectra of negative samples (non-barium
stars) were randomly selected from LAMOST DR9 F, G, and K type
giants catalog and based on the following selection criteria: S/N
\textgreater 30 pixel$^-1$ (both on the g-band and r-band, which aim
to be consistent with the screening of positive samples according to
the rules of N19) and log $g$ \textless 3.5 (Giants are defined by
this criterion \citep{2014ApJ...790..110L}). Figure \ref{fig:fig3}
presents the stellar parameter space of our sample in the plane of
$T_\mathrm{eff}$-$\log g$.

The spectra were preprocessed based on the following steps.

\noindent(1) Correct the wavelength by the radial velocity: the
wavelengths were corrected by the following formula.
\begin{equation} w_{new}(n)=w(n)/(1+z), \end{equation} where $w(n)$
represents the wavelengths of an observed spectrum, and $z$
represents the redshift of this star. The $z$ was calculated by the
LAMOST 1D pipeline \citep{2001ChJAA...1..563L}.

\noindent(2) Normalization: The flux of each spectrum is normalized
to the range [0,1] according to the following formula.
\begin{equation} \hat{x}=\frac{x-min(x)}{max(x)-min(x)},
\end{equation} where $x$ represents flux of an observed spectrum,
and $min(x)$ and $max(x)$ represent the minimum and maximum flux of
the spectrum $x$.

% ---------------------------------------------------------------------------
\section{Method}
\label{sec:method}
% ---------------------------------------------------------------------------

\subsection{ LGBM Algorithm }

\begin{figure*} \centering \includegraphics[width=18cm]{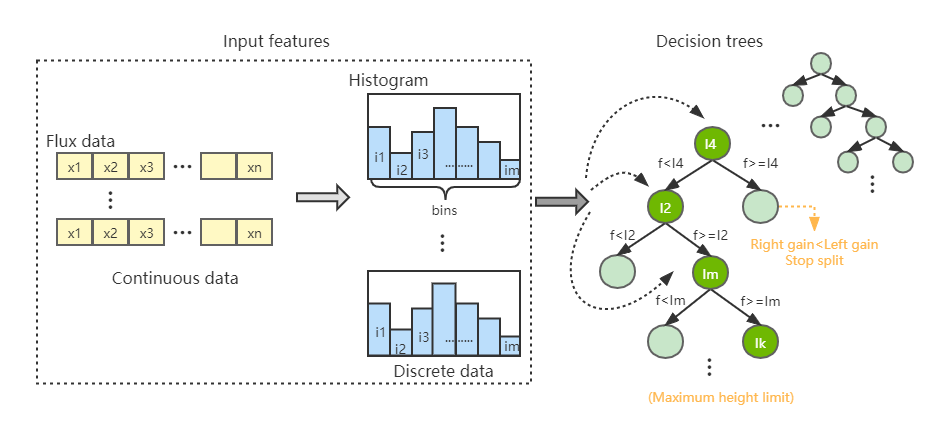}
\caption{This is a process diagram of how LGBM constructs a decision
tree for classification or prediction. The model discretizes the
input features and establishes the histogram. Then according to
these established histograms, features are selected as nodes to
split and build a decision tree. Each decision tree would give a
possible result, and the final result will be calculated according
to the results of all the decision trees. }\label{fig:lgbm}
\end{figure*}

The Light Gradient Boosting Machine algorithm proposed by
Microsoft in 2017 is a gradient boosting tree framework, which is an
ensemble learning algorithm \citep{ke2017lightgbm}. Figure
\ref{fig:lgbm} shows the theoretical process of the construction of
an LGBM model. An LGBM model consists of several decision
trees. Each tree is a weak learner, which is built by splitting leaf
nodes. LGBM uses the histogram method to split features and build
leaf nodes. As features are input into the model, continuous data
are discretized and histograms are built. Each unit of the histogram
consists of statistically discrete data, which are called a bin. For
each feature, the histogram stores two kinds of information: the sum
of gradients for the samples in each bin ($H[i].g$), and the number
of samples in each bin ($H[i].n$). The model traverses histograms
instead of the whole data and splits features according to the bin
with the maximum gain. The steps for calculating gain are as
follows.

The first is calculating gradients. Suppose that $H$ denotes a
storage structure in the histogram. For the ith bin in the
histogram, the current gradient sum of the left node is
\begin{equation} S_L=S_L+H[i].g; \end{equation}

H[i].g represents the gradient value of the corresponding feature of
each bin. The number of samples on the left leaves is
\begin{equation} n_L=n_L+H[i].n; \end{equation}

H[i].n represents the number of features corresponding to each bin.
For the input current gradient sum of parent node $S_P$, the output
current gradient sum of the right node is \begin{equation}
S_R=S_P-S_L; \end{equation}

Input the number of samples on parent leaves $n_P$ and calculate the
number of samples on right leaves \begin{equation} n_R=n_P-n_L;
\end{equation}

Then, the output loss function is
\begin{equation}
\Delta loss=\frac{S^2_L}{n_L}+\frac{S^2_R}{n_R}-\frac{S^2_P}{n_P}.
\end{equation}

The number of bins in the histogram is smaller than the size
of the continuous feature, which can reduce memory consumption. In
addition, the tree keeps growing until the maximum depth limit is
reached, which can prevent overfitting.

Iteration method assures the minimum loss of the loss
function when building a decision tree. The result of the model is a
comprehensive consideration of all weak learners' results: combining
them into a strong learner. The results are held in the leaves
which are at the ends of the branches. In the classifier, a result is a
number that reflects the degree of being a barium star or not. In
prediction, the result reflects the most probable value. When the
depth of trees reaches the set maximum depth (max\_depth), the
process of building the LGBM model will stop. In the testing
process, after inputting sample sets into the built model, every
sample starts from the root node and goes through the branches down
to the leaves, and then the results can be obtained.

The key parameters of the LGBM algorithm which we used are as
follows. \begin{itemize} \item n\_estimators, which define the
number of decision trees (the number of iterations), which
is the first parameter to be considered. Too small or too large will
lead to underfitting or overfitting. \item max\_depth, which
defines the maximum depth of each decision tree. This parameter is
inversely proportional to overfitting. \item num\_leaves, which
defines the number of leaves on each decision tree. It controls the
complexity of the tree model. Ideally, num\_leaves $\le
2^{max\_depth}$. \end{itemize} For the LGBM models, we set the estimators from 100 to 1000 and the step size is 100. After training, we found that the model gradually fitted after about 300 for the Sr classifier, and 200 for Ba classifier. These fields then were narrowed down to 200 to 400 for Sr and 100 to 300 for Ba, both in steps of 50. After continually training and searching for fitting
points, the fields were narrowed eventually to 300 to 310 for Sr and
215 to 225 for Ba, both in steps of 1. In this way, we can be sure
that estimators should be set at 303 for Sr classifier and 220 for
Ba classifier. The max depth (from 3 to 14 in steps of 1 initially)
and leaves' number (from 10 to 30 in steps of 1 initially) are set
in a similar way. Besides, we add hyperparameters (the minimum
weight of child nodes and the minimum number of samples, L1 and L2
regularization coefficient, etc.) to prevent over-fitting. The
procedure of parameters searching is supported by GridSearchCV,
which is a tool provided by scikit-learn \citep{abraham2014machine}
and is used in both classification and prediction to set
parameters.

\subsection{SVM}

SVM is a two-class classification model which is based on
statistical learning theory. SVM was firstly introduced by
\citep{1995The} and then was widely employed to solve astronomical
problems
\citep{2008A&A...478..971H,10.1111/j.1365-2966.2012.21191.x,
BU201435}. Here is a brief introduction to the principle of the SVM
algorithm.

We input the training data as
$T={(x_1,y_1),(x_2,y_2),...,(x_N,y_N)}$, where $x_i \in
R^n, y_i \in \{-1,1\}, i=1,2,...,N.$ , $x_i$ represents
feature vectors and $y_i$ represents category markers. We suppose
there is a hyperplane that separates positive examples from negative
ones. The points x that lie on the hyperplane satisfy $\omega \cdot
x+b=0$, where $\omega$ is normal to the hyperplane, and $|b|$ is the
perpendicular distance from the hyperplane to the origin. Then SVM
can be formulated as:
    \begin{equation}
        \left\{
        \begin{array}{lr}
            max\frac{1}{\|\omega\|}+C\sum_{i=1}^N\epsilon, & \\
            subject\quad to\quad y_i(\omega^Tx_i+b)\geq1-\epsilon _i, i=1,2,...,N,& \\
            \epsilon _i\geq0, i=1,2,...,M,
        \end{array}
        \right.
    \end{equation}
where $C$ is the penalty factor and $\epsilon _i$ are the slack
variables. Then this problem can be transformed into the following
formula \begin{equation}
    min_\alpha \frac{1}{2} \sum_{i=1}^N \sum_{j=1}^N \alpha _i\alpha _jy_iy_j(x_i \cdot x_j)-\sum_{i=1}^N\alpha _i, \label{con:eqSVM0}
\end{equation}
which is subject to
\begin{equation}
    \left\{
    \begin{array}{lr}
        \sum_{i=1}^N\alpha_iy_i=0, & \\
        0\leq \alpha _i \leq C, & i=1,2,...,N,
    \end{array}
    \right.
\end{equation} where $\alpha_i$ are the Lagrange multipliers
$\alpha$ for each sample ($x_i,y_i$). After using sequential minimal
optimization (SMO) to find the unique variable $\alpha _i^*$, in the
target function, we can obtain the optimal results. In this case,
$\omega$ is
    \begin{equation}
        \omega ^*=\sum_{i=1}^N \alpha _i^*y_ix_i.
\end{equation}

When the features are nonlinear. % as in Figure \ref{fig:svm},
SVM introduces the kernel function $K(x,z)$ to map the data set into
high-dimensional space and classifies it in this space by a linear
classifier. The equation (\ref{con:eqSVM0}) can be replaced by the
following formula \begin{equation}
    min_\alpha \frac{1}{2} \sum_{i=1}^N \sum_{j=1}^N \alpha _i\alpha _jy_iy_jK(x_i \cdot x_j)-\sum_{i=1}^N\alpha _i.
\end{equation} There are different types of kernel functions that
can be selected. In our work, we used a radial basis
function which is the most widely used and has good performance in
both large and small samples: $K(x,z)=exp(-\frac{\|x-z\|^2}{2\sigma
^2})$, where $z$ is the kernel function center and $\sigma $ is the
width parameter of the function that controls the radial scope of
the function. In this case, $\omega$ is
    \begin{equation}
        \omega ^*=\sum_{i=1}^N\alpha _i^*y_iK(x_i \cdot x_j).
    \end{equation}
    Then we just need to compute the sign of the following function
    \begin{equation}
        f(x)=sign(\omega \cdot x+b).
\end{equation} 
%\begin{figure} \centering

%\includegraphics[width=7cm]{svm.eps} \caption{SVM classification
%uses the kernel trick. A kernel function transforms nonlinear
%features into linearly separable features by mapping them to a
%higher dimensional linearly separable space.}\label{fig:svm}
%\end{figure}

% ---------------------------------------------------------------------------
\section{Experiment}
\label{sec:experiment}

% ---------------------------------------------------------------------------

There are mainly two research objectives for this
experiment. First, train the classification model using the barium
samples (L18 and N19) to separate the barium stars from the normal
giant stars. This model will be used to search for new barium giant
candidates from LAMOST DR9 in subsequent work. Second, building the
element abundance prediction model using the barium samples with
[Ba/Fe] and [Sr/Fe] labels(N19). This model will be used to estimate
[Ba/Fe] and [Sr/Fe] of the newly found barium giant stars in the
subsequent work.
The following content of this chapter will introduce our
experiment process in detail.

\subsection{ Performance Metric }\label{4.1}

The performance of a machine-learning classification model
is usually evaluated by precision, recall, and F1-score
\citep{zbMATH02118140}, which are parameters calculated from the
confusion matrix. These three evaluation criteria are defined as
follows. \begin{equation} precision=\frac{TP}{TP+FP};
\end{equation} \begin{equation} recall=\frac{TP}{TP+FN};
\end{equation} \begin{equation}
F1-score=\frac{2*(Precision*Recall)}{(Precision+Recall)}.
\end{equation} Since the test set consists of labeled
samples, it is easy to judge if the classifier's classification
results are correct on the test set. TP is the number of true barium
samples that are correctly classified as positive samples by the
model. FP is the number of non-barium stars that are misclassified
as positive samples by the model. Similarly, TN is the number of
non-barium stars that are correctly classified as negative samples
by the model, and FN is the number of true barium stars that are
misclassified as negative stars by the model.

Precision is defined as the percentage of correct barium
star predictions of all stars classified as positive, while the
recall is the fraction of barium stars correctly classified as
positive to the total number of barium samples.  F1-score is the
harmonic mean of the precision and recall.

In prediction, two methods are used to evaluate the performance of
the model.

\noindent1.Mean absolute error (MAE). It can avoid
compensating error and accurately reflect the actual prediction
error, which is a commonly used performance metric of a regression
model.
\begin{equation}MAE=\frac{1}{n}\sum_{i=1}^n\\y_{test}^i-Y_{test}^i,
\end{equation} where $n$ is the number of samples in the testing
data set, $y_{test}^i$ is the target value of samples and
$Y_{test}^i$ is the result of the predictor corresponding to
$y_{test}^i$. It can measure the actual error between the target
value and the result of the predictor.

\noindent2.Standard deviation ($\sigma$). \begin{equation} \sigma =
\sqrt{\frac{\sum_{i=1}^n(E^i-E^{i'})^2}{n}}, \end{equation} where
$E^i=y_{test}^i-Y_{test}^i$, and $E^{i'2}$ is the average of $E^i$.
It can measure the dispersion degree of the difference between the
prediction result and the target value, and evaluate the stability
of the LGBM model.

\subsection{ Input Feature Selection}\label{4.2}

The method to distinguish barium stars mainly depends on the
spectral line features. For example, N19 identified the
s-process-rich candidates by comparing the strengths of the most
conspicuous Sr\,{\sc ii} ( 4077 {\AA} and 4215 {\AA}) and Ba\,{\sc
ii} lines( 4554 {\AA} and 4934 {\AA}) in template with observed
spectra. L18 distinguished their barium samples based on the strong
lines of s-process elements, particularly Ba\,{\sc ii} at 4554 {\AA}
and Sr\,{\sc ii} at 4077 {\AA}. Considering our small sample size,
we should choose the same feature band mentioned above as the input
feature instead of the whole spectrum which contains too much
irrelevant information. Well-chosen input features can improve
classification accuracy substantially, or equivalently, reduce the
amount of training data needed to obtain the desired level of
performance \citep{2002Forman}. However, an important fact
has to be considered. Barium stars are usually enriched not only
with Ba or Sr but also with other s-process elements, such as Y, Zr,
La, Ce, Nd, etc \citep{2007Smiljanic,2018K}. Although quite a lot of
spectral lines of these elements are weak and it is not easy to
estimate the abundances of corresponding elements accurately in
low-resolution spectra, these characteristics can still be used as
an important reference for the barium star criterion. Therefore,
this study adopts two feature input methods to compare the effect
and efficiency.

\begin{figure*}
\centering
\subfigure{\includegraphics[width=17cm]{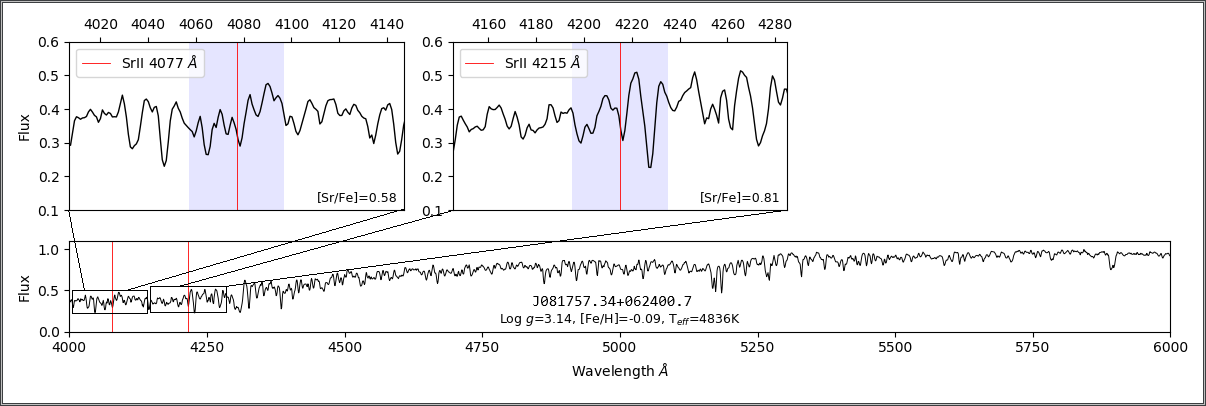}}
\centerline{\footnotesize Sr enhanced}
\\
\centering \subfigure{\includegraphics[width=17cm]{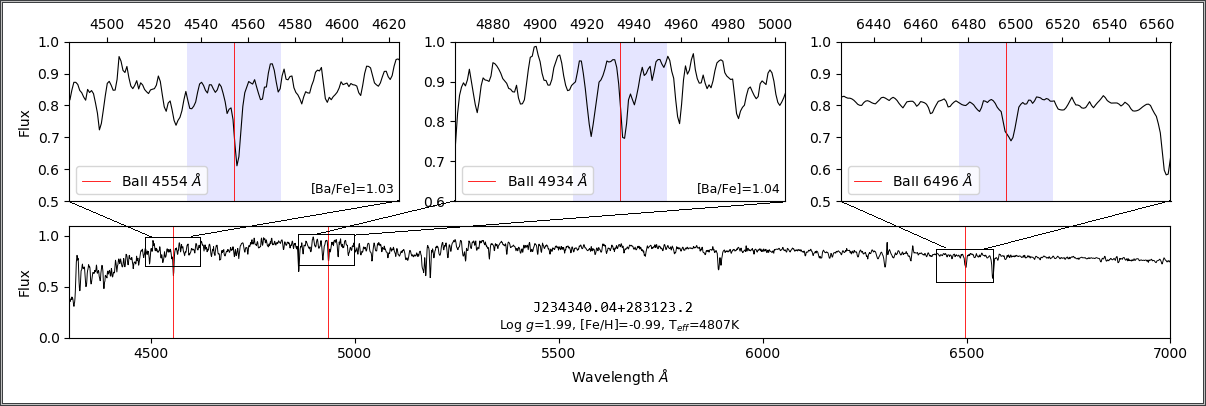}}
\centerline{\footnotesize Ba enhanced} \caption{ Local continuum
spectra for Sr-only (J081757.34+062400.7) and Ba-only
(J234340.04+283123.2) candidates randomly selected from the N19
dataset. The zoom-in ranges of wavelengths are feature bands of 70
{\AA}, and the light blue part is feature bands of 20 {\AA}.
\label{fig:feature}} \end{figure*}

The first is to select several spectral bands containing
the absorption line adopted by N19 and L18 as the input feature. We
have additionally adopted the Ba\,{\sc ii} line at 6496 {\AA}. This
absorption line is prominent and often used as an important basis
for the discrimination of barium stars. When selecting the feature
bandwidth on each side of an absorption line, we compared two
wavelength regions: 70 Å and 20 Å. These selected spectral bands
are shown in Figure \ref{fig:feature}. The second is inputting the
entire spectrum as an input feature, and feature selection depends
entirely on machine learning algorithms. Using this method, the
absorption line features of other s-process elements will be a
useful supplement although useless interference information is also
increased.

\subsection{ Classification Construction}
As mentioned in the introduction, N19 divided the
s-process-rich candidates they found into three types: Sr-only,
Ba-only, and both Sr- and Ba- enhancement candidates.
\cite{2021Karinkuzhi} analyzed 15 of the brightest targets of the
s-process-rich candidates searched by N19 based on high-resolution
spectra, which consist of 13 Sr-only stars and two Ba-only stars.
The analysis results show that four Sr-only stars present no
s-process overabundances. They considered that the principal reason
for the misclassification is that the Sr\,{\sc ii} lines at 4077 and
4215 {\AA} used by N19 are easily saturated and the 4215 {\AA} line
is strongly influenced by the strong CN bandhead at $\lambda$ 4216
{\AA}. In fact, three of the four no-s stars considered as Sr-only
by N19 are those being N-rich. \cite{2021Karinkuzhi} compared their
derived Ba abundances with those determined by N19 based on the
Ba\,{\sc ii} lines at 4554 and 4934 {\AA} and found that the
agreement is much better for Ba than for Sr. Although our purpose is
to search for s-process-rich candidates containing Sr-rich or
Ba-rich stars, based on the reasons above and our preliminary
experiment, instead of taking Sr-only and Ba-only as the same
category, we have built two classifiers to identify Sr-rich and
Ba-rich candidates respectively, which are called Sr-classifier and
Ba-classifier. 

We divided the positive sample into two parts. The first part
consists of 1629 Sr-enhanced candidates, which includes 819
Sr-enhanced (642 Sr-only and 177 both Sr- and Ba-) barium stars from
N19 and 810 barium stars from L18. The other part consists of 1035
Ba-enhanced candidates, which includes 225 Ba-enhanced (48 Ba-only
and 177 both Sr- and Ba-) barium stars from N19 and 810 barium stars
from L18. Two independent classifiers are constructed based on the
two parts of the data set respectively.

\cite{2020Time} used interpolation to augment the
time-series data. This method interpolates the virtual new
collection points into each collection point of the original
time-series, which obtains the interpolated time-series. Then they
generate a new time-series by extracting random data points. Data
augmentation can be achieved based on this method and can keep the
trend information of the original time-series. Considering the
similar 2D data patterns, we used the same method on spectral data
to improve the generalization capabilities of the machine learning
model. The basic idea is to interpolate the information of flux at
the wavelengths of every spectrum with the range of 4000-8098 {\AA},
and then extract data points at regular intervals to generate new
spectra. The number ratio of the training set and testing
set is 8:2 by random division. We enhanced the training data set of Sr-enhanced up to 3893 and Ba-enhanced up to 2474 by this method, but the testing sets were not
augmented. Table \ref{tab4-2} lists the composition of the data
set.

\begin{table*} \caption{The Comparison Results of LGBM and SVM with
Different Input Features. Checkmarks mean the feature bands
selected by the model and the corresponding feature width
intercepted.}
    \centering
    \begin{tabular}{cccccccc}
        \hline
        \hline
        \multicolumn{2}{c}{Classifiers} & \multicolumn{3}{c}{Sr enhanced}& \multicolumn{3}{c}{Ba enhanced}\\
        \hline
        \multirow{2}*{Feature width}& 70\AA&\checkmark & & &\checkmark & & \\
        & 20\AA & &\checkmark& & &\checkmark& \\
        \cline{1-8}
        \multirow{6}*{Feature bands}& 4077\AA (Sr II)&\checkmark&\checkmark& & & & \\
        & 4215 \AA (Sr II)&\checkmark&\checkmark& & & & \\
        & 4554 \AA (Ba II)& & & &\checkmark&\checkmark& \\
        & 4934 \AA (Ba II)& & & &\checkmark&\checkmark& \\
        & 6496 \AA (Ba II)& & & &\checkmark&\checkmark& \\
        \cline{1-8}
        Whole spectrum & 4000-8000\AA& & &\checkmark& & &\checkmark \\
        \hline
        \multirow{3}*{LGBM}&F1-score&94.60\% &95.07\% & 96.92\% & 96.70\% & 96.71\% & 96.87\% \\
        & Recall & 92.82\% & 93.54\% & 96.05\% & 96.84\% & 97.13\% & 97.70\% \\
        & Precision& 96.46\%& 96.66\% & 97.81\% & 96.56\% & 96.30\% & 96.03\%\\
        \hline
        \multirow{3}*{SVM}&F1-score&93.51\% &91.93\% & 96.83\% & 96.47\% & 96.01\% & 96.02\%\\
        & Recall & 89.23\% & 86.89\% & 95.29\% & 96.55\% & 96.84\% & 95.83\% \\
        & Precision& 98.22\%& 97.58\% & 98.37\% & 96.39\% & 95.20\% & 96.21\% \\
        \hline
        \hline
    \end{tabular}
    \label{tab1}
\end{table*}

In this experiment, we employed the LGBM algorithm. Considering that
SVM has excellent performance on small samples, we also adopted the
SVM algorithm for comparison. The classification results of
inputting different feature bands are compared and shown in Table
\ref{tab1}. There is a point here. The absorption line
profiles of an element are affected not only by the abundance value
of the element but also by other factors, mainly atmospheric
parameters. For the feature bands input method, atmospheric
parameters ($T_{eff}$, Log $g$, [Fe/H]) have been normalized
to [0,1] and added as the feature input because the model cannot
determine them from several bands with the wavelength range of 20
{\AA} or 70 {\AA}.

It should be noted that the results in Table \ref{tab1} are
the average of three metrics of five-fold. In order to evaluate the
reliability of our model, the five-fold cross-validation was used
and the Sr-enhanced samples and Ba-enhanced samples were divided
into five equal parts respectively. One of five parts of all data
was taken as the test data, 10\% of the remaining four parts was
used as a validation set, and the rest was used as a training set.
The process was repeated five times by moving the test data portion.

For the overall performance of the classification in Table
\ref{tab1}, the LGBM algorithm performs excellently and stably in
general. For the Sr classifier, the different selection of features has a greater
impact on the Recall, especially for the SVM algorithm. For the Ba classifier,
there is no significant difference between the classification results obtained
 by the two algorithms using different feature bands.
In other words, using prior knowledge to select different input features does not improve 
the performance of the classifier, the reason may be that when we use the whole spectra (4000-8000 {\AA}) as input feature, 
the other s-process elements we analyzed in \ref{4.2} above, such as Y, Zr,
La, Ce and Nd, provide additional judgment bases, which can overcome
the interference caused by other absorption lines.

In this experiment, the best model (The feature selection is the whole spectra): Sr classifier consists of 303 decision trees (i.e.,
$n\_estimatiors=303$) and 14 leaves on each tree (i.e., $num\_leaves
= 14$) with heights of 12 (i.e., $max\_depth=12$) based on LGBM
model. Ba classifier based on LGBM consists of 220 decision trees
(i.e., $n\_estimatiors = 220$) and 9 leaves on each tree (i.e.,
$num\_leaves = 9$) with heights of 4 (i.e., $max\_dept =4$). The
parameters based on SVM are as follows: the penalty parameter is
759 (i.e., $C = 759$) and kernel coefficient is 0.005 (i.e.,
$gamma = 0.005$) in Sr classifier, and the penalty parameter is
1370 (i.e., $C = 1370$) and kernel coefficient is 0.0005 (i.e.,
$gamma = 0.0005$) in Ba classifier. The feature selection is 20 {\AA} band: Sr
classifier consists of 523 decision trees (i.e., $n\_estimatiors =
523$) and 13 leaves on each tree (i.e., $num\_leaves = 13$) with
heights of 5 (i.e., $max\_depth = 5$) based on LGBM model. Ba
classifier based on LGBM consists of 217 decision trees (i.e.,
$n\_estimatiors = 217$) and 18 leaves on each tree (i.e.,
$num\_leaves = 18$) with heights of 5 (i.e., $max\_depth = 5$).
The parameters based on SVM are as follows: the penalty parameter is
20 (i.e., $C = 20$) and kernel coefficient is 15 (i.e.,
$gamma = 15$) in Sr classifier, and the penalty parameter is
1848 (i.e., $C = 1848$) and kernel coefficient is 0.5 (i.e.,
$gamma = 0.5$) in Ba classifier. The feature selection is 70 {\AA} band: Sr
classifier consists of 665 decision trees (i.e., $n\_estimatiors =
665$) and 13 leaves on each tree (i.e., $num\_leaves = 13$) with
heights of 4 (i.e., $max\_depth = 4$) based on LGBM model. Ba
classifier based on LGBM consists of 211 decision trees (i.e.,
$n\_estimatiors = 211$) and 9 leaves on each tree (i.e.,
$num\_leaves = 9$) with heights of 5 (i.e., $max\_depth = 5$).
The parameters based on SVM are as follows: the penalty parameter is
30 (i.e., $C = 30$) and kernel coefficient is 1 (i.e.,
$gamma = 1$) in Sr classifier, and the penalty parameter is
1378 (i.e., $C = 1378$) and kernel coefficient is 0.01 (i.e.,
$gamma = 0.01$) in Ba classifier.

\begin{table*}
\caption{Abundance of Partial Barium Giants Reported by N19.}
\centering
\begin{tabular}{cccccccc}
\hline
\hline
\multirow{2}*{2MASS}&\multicolumn{2}{c}{[Sr/Fe]}&\multicolumn{2}{c}{[Ba/Fe]}&\multirow{2}*{Sr-only}&\multirow{2}*{Ba-only}&\multirow{2}*{Sr- and Ba-enhancement}\\
\cline{2-5}
&4077{\AA}&4215{\AA}&4554{\AA}&4934{\AA}&&&\\
\hline
J231231.643+430416.61 &0.98 &0.98 &$-0.04$&$-0.06$&\checkmark&$\times$&$\times$ \\
J005748.459+425733.24 &0.92 &0.87 &0.39 &$-0.14$&\checkmark&$\times$&$\times$ \\
J060926.438+283923.42 &0.76 &$-0.13$&$-0.01$&$-0.20$&\checkmark&$\times$&$\times$ \\
J231517.618+164636.08 &$-0.16$&$-0.07$&1.04 &1.12&$\times$&\checkmark&$\times$\\
J032435.984+36086.22 &0.62 &$-0.04$&1.00 &0.93&$\times$&\checkmark&$\times$ \\
J183447.556+354651.68 &0.75 &0.73 &0.85 &1.04&$\times$&\checkmark&$\times$ \\
J064614.085+162156.82 &0.93 &0.88 &1.01 &0.86&$\times$&$\times$&\checkmark\\
J222439.264+083051.67 &0.80 &0.53 &0.84 &$-0.11$&$\times$&$\times$&\checkmark \\
J024832.227+372351.13 &0.83 &$-0.05$&1.02 &0.89&$\times$&$\times$&\checkmark \\
\hline
\hline
\end{tabular}
\label{tab4-1}
\end{table*}

\begin{table*} \caption{Number of Training and Testing Data Sets in
the Classification.} \centering \begin{tabular}{cccccccc} \hline
\hline
\multicolumn{2}{c}{Numbers} & \multicolumn{3}{c}{Sr classifier}& \multicolumn{3}{c}{Ba classifier}\\
\hline
\multicolumn{2}{c}{\multirow{3}{*}{Spectra number of barium stars}}&\multirow{3}{*}{1629}&\multirow{2}{*}{819 (N19)}&642 Sr-only&\multirow{3}{*}{1035}&\multirow{2}{*}{225 (N19)}&48 Ba-only\\
\cline{5-5}
\cline{8-8}
&&&&177 Sr\&Ba&&&177 Sr\&Ba\\
\cline{4-5}
\cline{7-8}
&&&\multicolumn{2}{c}{810 (L18)}&&\multicolumn{2}{c}{810 (L18)}\\
\hline
\multirow{3}{*}{Positive samples number}&training set(before augmented)&\multicolumn{3}{c}{1303}&\multicolumn{3}{c}{828}\\
\cline{2-2}
&training set(after augmented)&\multicolumn{3}{c}{3893}&\multicolumn{3}{c}{2474}\\
\cline{2-2}
&testing set&\multicolumn{3}{c}{326}&\multicolumn{3}{c}{207}\\
\multicolumn{2}{c}{Ratios of positive and negative samples}&\multicolumn{3}{c}{1:1}&\multicolumn{3}{c}{1:1}\\
\hline
\hline
\end{tabular}
\label{tab4-2}
\end{table*}

\begin{figure*} \centering \includegraphics[width=17cm]{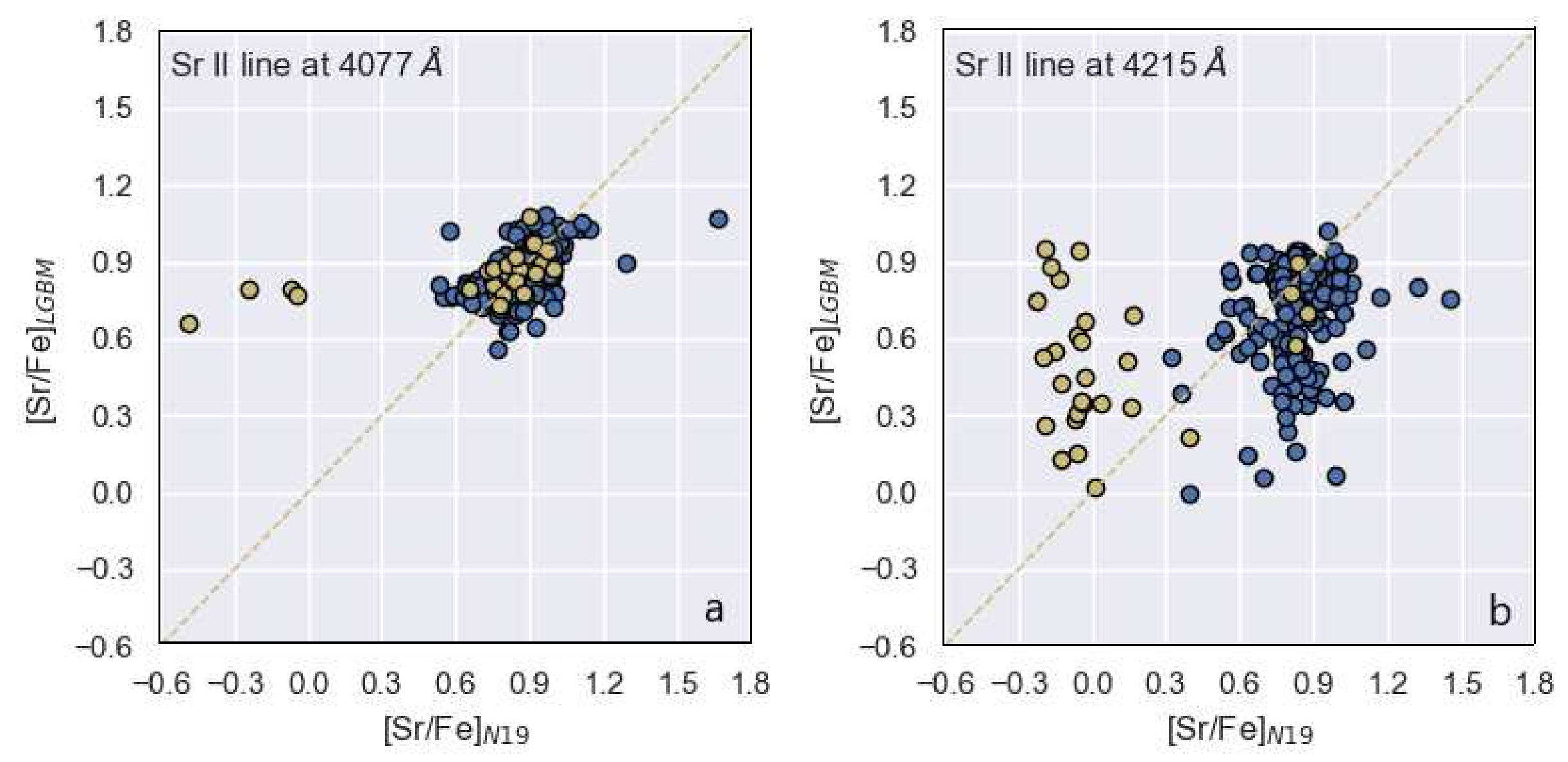}
\caption{Comparisons of the Sr abundances calculated by our work
([Sr/Fe]$_{LGBM}$) and values determined from N19 ([Sr/Fe]$_{N19}$).
The yellow line represents that the predicted value is equal to the
target value. The yellow dot represents the point at which
$\mid$[Sr/Fe$]_{4077 \AA}$ - [Sr/Fe$]_{4215 \AA}\mid \geq 0.5$ in
N19. The left shows the differences between [Sr/Fe]$_{LGBM}$ and
[Sr/Fe]$_{N19}$ in the prediction of 4077 {\AA} (Sr\,{\sc ii}), and
the right shows these differences in the prediction of 4215 {\AA}
(Sr\,{\sc ii}).}\label{fig:srpre} \end{figure*}

\begin{figure*} \centering \includegraphics[width=17cm]{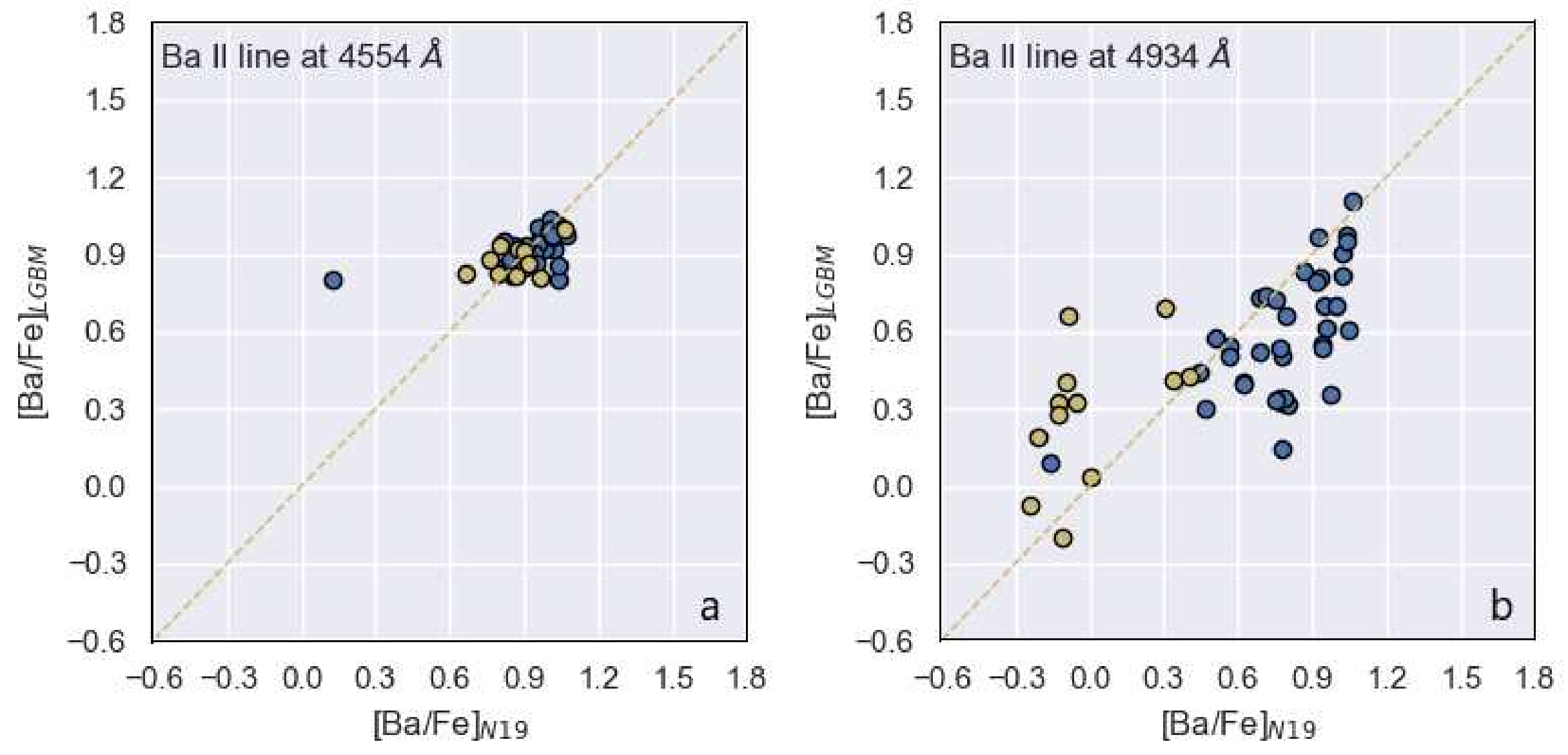}
\caption{Comparisons of the Ba abundances calculated by our work
([Ba/Fe]$_{LGBM}$) and values determined from N19 ([Ba/Fe]$_{N19}$).
The yellow line signifies that the predicted value is equal to the
target value. The yellow dot represents the point at which
$\mid$[Ba/Fe$]_{4554 \AA}$ - [Ba/Fe$]_{4934 \AA}\mid \geq 0.5$ in
N19. The left shows the differences between [Ba/Fe]$_{LGBM}$ and
[Ba/Fe]$_{N19}$ in the prediction of 4554 {\AA} (Ba\,{\sc ii}), and
the right shows these differences in the prediction of 4934 {\AA}
(Ba\,{\sc ii}).}\label{fig:bapre} \end{figure*}

\subsection{ Prediction Construction }

Since L18 did not provide [Ba/Fe] and [Sr/Fe] of barium stars, we
only used N19 as the data set. Table \ref{tab4-1} provides the
[Ba/Fe] and [Sr/Fe] at different lines of several barium stars that
come from N19.

In this work, we use the LGBM algorithm to predict the [Ba/Fe] and
[Sr/Fe]. Sr enhanced predictor consists of 800 trees (i.e.,
n\_estimatiors = 800) and 20 leaves on each tree (i.e., num\_leaves
= 20) with heights of 10 (i.e.,max\_depth = 10). We also set other
parameters (i.e., min\_child\_samples = 20, min\_child\_weight =
0.01).  When continuously adjusting parameters through GridSearchCV,
the parameters of the Ba-enhanced predictor were numerically the
same as Sr enhanced predictor. Using [Ba/Fe] and [Sr/Fe] provided by
N19 as labels, the selection of model feature bands and the
performance of our models are shown in Table \ref{tabpre}.

As seen from Table \ref{tabpre}, the prediction result of [Sr/Fe] at
the 4077 {\AA} is less scattered, MAE = 0.09 and the standard
deviation ($\sigma$) is about 0.16; while the prediction result of
[Ba/Fe] at 4554 {\AA} is closer to its label value than that at line
4934 {\AA}, with  MAE = 0.07 and $\sigma$ = 0.12.

Figure \ref{fig:srpre} and Figure \ref{fig:bapre} show the
comparison between the abundance provided by N19 and the LGBM
prediction results. From Figure \ref{fig:srpre}(b), it can be seen
that in the range of [Sr/Fe]$_{N19}$ $\textless$ 0.3, the prediction
value of [Sr/Fe]$_{LGBM}$ is much higher. The same situation also
appears in [Ba/Fe]$_{4934 \AA}$. In this part of the data (most of
the yellow dots), it can be seen from the lists provided by N19 that
the difference in abundance obtained by different absorption lines
of the same element is relatively large: $\mid$[Sr/Fe$]_{4077 \AA}$
- [Sr/Fe$]_{4215 \AA}\mid \geq 0.5$, $\mid$[Ba/Fe$]_{4554 \AA}$ -
[Ba/Fe$]_{4934 \AA}\mid \geq 0.5$. Therefore, the high dispersion of
prediction results in this part is probably due to the insufficient
accuracy of the sample labels.

\begin{table*}
\caption{The Results of LGBM in Predicting [Ba/Fe] and [Sr/Fe].}
\centering
\begin{tabular}{cccccc}
\hline
\hline
\multicolumn{2}{c}{\multirow{2}*{Predictors}}&\multicolumn{2}{c}{[Sr/Fe]}&\multicolumn{2}{c}{[Ba/Fe]} \\
\cline{3-6}
& & 4077{\AA} & 4215{\AA} & 4554{\AA} & 4934{\AA} \\
\hline
\multirow{4}*{Feature bands}& 4057-4097{\AA}(Sr\,{\sc ii}) & \checkmark & & &  \\
& 4195-4235{\AA}(Sr\,{\sc ii}) & & \checkmark & &  \\
\cline{2-6}
& 4534-4574{\AA}(Ba\,{\sc ii}) & & & \checkmark & \\
& 4925-4965{\AA}(Ba\,{\sc ii})& & & & \checkmark \\
\hline
\multirow{2}*{Performance}& MAE & 0.09 & 0.20 & 0.07 & 0.26\\
& $\sigma$ & 0.16 & 0.29 & 0.12 & 0.32\\
\hline
\hline
\end{tabular}
\label{tabpre}
\end{table*}

The high-resolution abundance analysis of 15 barium star candidates
of \cite{2021Karinkuzhi} showed that [Sr/Fe] obtained by N19 based
on low-resolution spectra is generally higher, while [Ba/Fe] is
generally lower. This means that searching for barium stars based on
the model of the Ba classifier will get more reliable barium star
candidates than the Sr classifier, while an accurate abundance value
will depend on high-resolution spectral analysis.

\subsection{ Comparison with Other Methods }

\begin{table*} \caption{The Comparison Results of SVM, KNN, RF and
XGBoost based on the whole spectra.} \centering
\begin{tabular}{ccccccc} \hline \hline
\multirow{2}{*}{Model}& \multicolumn{3}{c}{Sr enhanced}& \multicolumn{3}{c}{Ba enhanced}\\
\cline{2-7}
&F1-score& Recall&Precision&F1-score & Recall&Precision\\
\hline
KNN&92.58\%& 94.08\%&91.13\%&91.87\%&95.69\%&88.34\%\\
RF&94.00\%& 92.82\%&95.21\%&94.68\%&94.54\%&94.81\%\\
XGBoost&96.90\%& 96.18\%&97.63\%&96.29\%&96.84\%&95.74\%\\
SVM&96.83\%& 95.29\%&98.37\%&96.02\% &95.83\%&96.21\%\\
LGBM&96.92\%& 96.05\%&97.81\%&96.87\%&97.70\%&96.03\%\\
\hline
\hline
\end{tabular}
\label{tab3}
\end{table*}

In addition to LGBM and SVM, we also adopt three other popular
algorithms (KNN, Random Forest (RF) and XGBoost) to compare the
classifier performance in the whole feature bands, which often
perform well in data science tasks. RF and XGBoost are similar to
LGBM which is based on decision trees. For the Sr
classification model, we set the decision tree number to 297 and the
maximum depth to 12 in RF, while we set the decision tree number to
401, the maximum depth to 7 and the minimum weight of child nodes
to 5 in XGBoost. For the Ba classification model, we set the
decision tree number to 735 and the maximum depth to 13 in RF, while we
set the decision tree number to 260, the maximum depth to 5 and the
minimum weight of child nodes to 5 in XGBoost. KNN is a classical
machine learning algorithm that classifies barium stars by measuring
the distance between different features. We set 18 neighbors near
each sample to find for measurement both in Sr classifier and Ba classifier. Through repeated training and testing,
we finally arrived at the result of the comparison, which is shown
in Table \ref{tab3}. We can see that the LGBM still performs best in
general, and XGBoos has a very close excellent performance.

% ---------------------------------------------------------------------------
\section{Conclusion}
\label{sec:conclusion}
% ---------------------------------------------------------------------------

We constructed an Sr classifier, Ba classifier and abundance
prediction models based on small samples of barium star candidates.
SVM, KNN, RF, XGBoost and LGBM are applied for comparison in
classifiers. The results show that the LGBM algorithm performs best
on identifying barium stars, for Sr classification,
Precision=97.81\%, Recall=96.05\%; for Ba classification,
Precision= 96.03\%, Recall=97.70\%. The prediction results show that
Sr predictor based on Sr\,{\sc ii} at 4077 {\AA} and Ba predictor
based on Ba\,{\sc ii} at 4554 {\AA} performed better.

Besides the powerful learning ability of machine learning,
the good classification results may also be related to our samples.
The positive samples we adopted all have prominent Ba or Sr
absorption lines, which is obviously different from normal giants.
For the prediction model, the predicted [Ba/Fe] at 4544{\AA} and
[Sr/Fe] at 4077 {\AA} are well consistent with the labels, which may
also be because the distribution range of label values is narrow,
and the predicted values of the model tend to fall into this range.

After the comparison from using different feature bands plus
atmospheric parameters and inputting the entire spectrum for the
training data, the results show that the precision and recall of the
entire spectrum are the best. This indicates that machine learning
algorithms are fully capable of learning useful features from
complex data to optimize their model parameters, even if the number
of training samples is not very large.

The results of the high-resolution spectral analysis show that
[Sr/Fe] of the data set from N19 is higher and [Ba/Fe] of most data
is lower. Therefore, the candidates obtained by Ba classifier will
be more reliable when using the model to search for barium star
candidates in the future.

This work is supported by the National Natural Science Foundation of
China (NSFC) under Grant Nos. 11803016, U1931209 and 11873037.
Software:
LGBM(\url{https://lightgbm.readthedocs.io/en/latest/pythonapi/lightgbm.LGBMClassifier.html}),
Scikit-learn: Machine Learning in Python
(\url{https://scikit-learn.org/stable/index.html}).

%--------------------------------------------------------------------------------------------
%--------------------------------------------------------------------------------------------

%***

\bibliography{main_edited}
\bibliographystyle{aasjournal}

\end{document}